\input harvmac
\input epsf

\def\hb{{1\over 2}R}

\input labeldefs.tmp

\writedefs


\Title{ \vbox{\baselineskip12pt \rightline{PUPT-2106}
 \rightline{hep-th/0401025}
 \vskip -.25in} }{ \vbox{
\centerline{Five-dimensional Gauge Theories and}
\vskip 0.1in
\centerline{ Unitary Matrix Models } } }

\centerline{ Martijn Wijnholt}
\vskip .1in
\centerline{\sl Department of Physics, Jadwin Hall, Princeton
University}
\centerline{\sl Princeton, NJ 08544, USA}
\centerline{\it  }

\vskip .1in

\vskip 0.3in

\centerline{\bf Abstract}

\noindent

The matrix model computations of effective superpotential terms in ${\cal N}=1$
supersymmetric gauge theories in four dimensions have been proposed to apply more
generally to gauge theories in higher dimensions.
We discuss aspects of five-dimensional gauge theory compactified on a circle, which leads to a unitary matrix model.

\Date{January 2004}


\lref\ArkaniHamedTB{
N.~Arkani-Hamed, T.~Gregoire and J.~Wacker,
``Higher dimensional supersymmetry in 4D superspace,''
JHEP {\bf 0203}, 055 (2002)
[arXiv:hep-th/0101233].
}

\lref\DijkgraafXK{
R.~Dijkgraaf and C.~Vafa,
``N = 1 supersymmetry, deconstruction, and bosonic gauge theories,''
arXiv:hep-th/0302011.
}

\lref\DV{
R.~Dijkgraaf and C.~Vafa,
``Matrix models, topological strings, and supersymmetric gauge theories,''
Nucl.\ Phys.\ B {\bf 644}, 3 (2002)
[arXiv:hep-th/0206255].
\hfill\break
R.~Dijkgraaf and C.~Vafa,
``On geometry and matrix models,''
Nucl.\ Phys.\ B {\bf 644}, 21 (2002)
[arXiv:hep-th/0207106].
\hfill\break
R.~Dijkgraaf and C.~Vafa,
``A perturbative window into non-perturbative physics,''
arXiv:hep-th/0208048.
}

\lref\NaculichHR{
S.~G.~Naculich, H.~J.~Schnitzer and N.~Wyllard,
``Matrix model approach to the N = 2 U(N) gauge theory with matter in the  fundamental representation,''
JHEP {\bf 0301}, 015 (2003)
[arXiv:hep-th/0211254].
\hfill\break
S.~G.~Naculich, H.~J.~Schnitzer and N.~Wyllard,
``The N = 2 U(N) gauge theory prepotential and periods from a  perturbative matrix model calculation,''
Nucl.\ Phys.\ B {\bf 651}, 106 (2003)
[arXiv:hep-th/0211123].
}

\lref\CachazoYC{
F.~Cachazo, N.~Seiberg and E.~Witten,
``Chiral Rings and Phases of Supersymmetric Gauge Theories,''
JHEP {\bf 0304}, 018 (2003)
[arXiv:hep-th/0303207].
}

\lref\CachazoZK{
F.~Cachazo, N.~Seiberg and E.~Witten,
``Phases of N = 1 supersymmetric gauge theories and matrices,''
JHEP {\bf 0302}, 042 (2003)
[arXiv:hep-th/0301006].
}

\lref\CachazoRY{
F.~Cachazo, M.~R.~Douglas, N.~Seiberg and E.~Witten,
``Chiral rings and anomalies in supersymmetric gauge theory,''
JHEP {\bf 0212}, 071 (2002)
[arXiv:hep-th/0211170].
}

\lref\CachazoPR{
F.~Cachazo and C.~Vafa,
``N = 1 and N = 2 geometry from fluxes,''
arXiv:hep-th/0206017.
}

\lref\CachazoJY{
F.~Cachazo, K.~A.~Intriligator and C.~Vafa,
``A large N duality via a geometric transition,''
Nucl.\ Phys.\ B {\bf 603}, 3 (2001)
[arXiv:hep-th/0103067].
}

\lref\KatzFH{
S.~Katz, A.~Klemm and C.~Vafa,
``Geometric engineering of quantum field theories,''
Nucl.\ Phys.\ B {\bf 497}, 173 (1997)
[arXiv:hep-th/9609239].
\hfill\break
S.~Katz, P.~Mayr and C.~Vafa,
``Mirror symmetry and exact solution of 4D N = 2 gauge theories. I,''
Adv.\ Theor.\ Math.\ Phys.\  {\bf 1}, 53 (1998)
[arXiv:hep-th/9706110].
}

\lref\IntriligatorPQ{
K.~A.~Intriligator, D.~R.~Morrison and N.~Seiberg,
``Five-dimensional supersymmetric gauge theories and degenerations of  Calabi-Yau spaces,''
Nucl.\ Phys.\ B {\bf 497}, 56 (1997)
[arXiv:hep-th/9702198].
}

\lref\GanorPC{
O.~J.~Ganor, D.~R.~Morrison and N.~Seiberg,
``Branes, Calabi-Yau spaces, and toroidal compactification of the N = 1  six-dimensional E(8) theory,''
Nucl.\ Phys.\ B {\bf 487}, 93 (1997)
[arXiv:hep-th/9610251].
}

\lref\IqbalEP{
A.~Iqbal and V.~S.~Kaplunovsky,
``Quantum deconstruction of a 5D SYM and its moduli space,''
arXiv:hep-th/0212098.
\hfill\break
A.~Iqbal and V.~S.~Kaplunovsky, To Appear.
}

\lref\AganagicWV{
M.~Aganagic, A.~Klemm, M.~Marino and C.~Vafa,
``Matrix model as a mirror of Chern-Simons theory,''
arXiv:hep-th/0211098.
}

\lref\AcharyaDZ{
B.~S.~Acharya and C.~Vafa,
``On domain walls of N = 1 supersymmetric Yang-Mills in four dimensions,''
arXiv:hep-th/0103011.
}

\lref\OoguriBV{
H.~Ooguri and C.~Vafa,
``Knot invariants and topological strings,''
Nucl.\ Phys.\ B {\bf 577}, 419 (2000)
[arXiv:hep-th/9912123].
}

\lref\LeungTW{
N.~C.~Leung and C.~Vafa,
``Branes and toric geometry,''
Adv.\ Theor.\ Math.\ Phys.\  {\bf 2}, 91 (1998)
[arXiv:hep-th/9711013].
}

\lref\AganagicQG{
M.~Aganagic, M.~Marino and C.~Vafa,
``All loop topological string amplitudes from Chern-Simons theory,''
arXiv:hep-th/0206164.
}

\lref\BrandhuberUA{
A.~Brandhuber, N.~Itzhaki, J.~Sonnenschein, S.~Theisen and S.~Yankielowicz,
``On the M-theory approach to (compactified) 5D field theories,''
Phys.\ Lett.\ B {\bf 415}, 127 (1997)
[arXiv:hep-th/9709010].
}

\lref\KannoVD{
H.~Kanno and Y.~Ohta,
``Picard-Fuchs equation and prepotential of five dimensional SUSY gauge  theory compactified on a circle,''
Nucl.\ Phys.\ B {\bf 530}, 73 (1998)
[arXiv:hep-th/9801036].
\hfill\break
T.~Eguchi and H.~Kanno,
``Five-dimensional gauge theories and local mirror symmetry,''
Nucl.\ Phys.\ B {\bf 586}, 331 (2000)
[arXiv:hep-th/0005008].
}

\lref\WittenQB{
E.~Witten,
``Phase Transitions In M-Theory And F-Theory,''
Nucl.\ Phys.\ B {\bf 471}, 195 (1996)
[arXiv:hep-th/9603150].
}

\lref\HollowoodCV{
T.~J.~Hollowood, A.~Iqbal and C.~Vafa,
``Matrix models, geometric engineering and elliptic genera,''
arXiv:hep-th/0310272.
}

\lref\HollowoodGR{
T.~J.~Hollowood,
``Five-dimensional gauge theories and quantum mechanical matrix models,''
JHEP {\bf 0303}, 039 (2003)
[arXiv:hep-th/0302165].
}

\lref\LawrenceJR{
A.~E.~Lawrence and N.~Nekrasov,
``Instanton sums and five-dimensional gauge theories,''
Nucl.\ Phys.\ B {\bf 513}, 239 (1998)
[arXiv:hep-th/9706025].
}

\lref\NekrasovCZ{
N.~Nekrasov,
``Five dimensional gauge theories and relativistic integrable systems,''
Nucl.\ Phys.\ B {\bf 531}, 323 (1998)
[arXiv:hep-th/9609219].
}

\lref\GanguliKR{
S.~Ganguli, O.~J.~Ganor and J.~A.~Gill,
``Twisted six dimensional gauge theories on tori, matrix models, and integrable systems,''
arXiv:hep-th/0311042.
}

\lref\KolFV{
B.~Kol,
``5d field theories and M theory,''
JHEP {\bf 9911}, 026 (1999)
[arXiv:hep-th/9705031].
}

\lref\AharonyBH{
O.~Aharony, A.~Hanany and B.~Kol,
``Webs of (p,q) 5-branes, five dimensional field theories and grid  diagrams,''
JHEP {\bf 9801}, 002 (1998)
[arXiv:hep-th/9710116].
}

\lref\VafaWI{
C.~Vafa,
``Superstrings and topological strings at large N,''
J.\ Math.\ Phys.\  {\bf 42}, 2798 (2001)
[arXiv:hep-th/0008142].
}

\lref\NekrasovRJ{
N.~Nekrasov and A.~Okounkov,
``Seiberg-Witten theory and random partitions,''
arXiv:hep-th/0306238.
}

\lref\KennawayJT{
K.~D.~Kennaway and N.~P.~Warner,
``Effective superpotentials, geometry and integrable systems,''
arXiv:hep-th/0312077.
}

\lref\AganagicXQ{
M.~Aganagic, K.~Intriligator, C.~Vafa and N.~P.~Warner,
``The glueball superpotential,''
arXiv:hep-th/0304271.
\hfill \break
K.~Intriligator, P.~Kraus, A.~V.~Ryzhov, M.~Shigemori and C.~Vafa,
 ``On low rank classical groups in string theory, gauge theory and matrix
models,''
arXiv:hep-th/0311181.
}

\newsec{Overview}

In the physics of four dimensional ${\cal N}=1$ supersymmetric gauge theories a special
r\^ole is played by the F-terms. The low energy effective superpotential terms
obey stringent constraints and can often be determined exactly.
Pushing the simplification to its extreme, it was argued in
\DV\ that for a large class of  examples the
computation can be done with only the planar diagrams in an auxiliary matrix model. The
canonical example is that of an ${\cal N}=2$ gauge theory deformed to
${\cal N}=1$ by a superpotential, where the planar diagrams can be summed up exactly.

In \DijkgraafXK\ it was suggested that matrix model methods can also be used
to analyse F-terms in  higher dimensional supersymmetric gauge theories.
Consider for instance a five dimensional $U(N)$ gauge theory with minimal ${\cal N} = 2$
SUSY. The matrix model methods of \DV\ apply to theories with only
four supercharges and isolated vacua
so we cannot tackle such a theory directly. However, suppose
we put the theory on a circle of radius
$R$ and coordinate $t$. The resulting theory has a Coulomb branch
parametrised by the eigenvalues of the complexified Wilson line
\eqn\aa{
U = P\exp(\int dt\, A_t + \varphi) .}
Now one may turn on a superpotential $W(U)$ with isolated critical
points. It also breaks the number of supercharges to four. This theory
exhibits gluino condensation and we can ask for the effective
glueball superpotential. In \DijkgraafXK\ it was proposed that the effective
superpotential is computed by a unitary matrix model. If one
includes hypermultiplets, the effective superpotential should be computed
by a matrix quantum mechanics.

As a test of these ideas, one may try to recover the
Seiberg-Witten curve for the (compactified) five-dimensional gauge
theory by turning off the superpotential. The curve has already been obtained through many other approaches, such as
integrable systems \NekrasovCZ, (p,q) five-branes \refs{\KolFV,\BrandhuberUA}, geometric engineering
\refs{\LawrenceJR,\KannoVD}, deconstruction \IqbalEP\ and instanton calculus \NekrasovRJ,
and the links between many of these calculations are known.
In section 2 we will review how the curve is obtained through
geometric engineering.

The routes to and through the relevant matrix model can be made to follow in
rough outline the routes that have been taken for the four-dimensional case.
We choose to focus on some of the earlier ideas, which emphasize the embedding
of the gauge theory in a larger string theory.  In
this spirit,
in section 3 we consider a D-brane setup for the five dimensional Yang-Mills theory
where the computation of F-terms can be argued to reduce to a matrix model.
The effective superpotential that the matrix model computes arises
from an open-closed duality in string theory where the D-branes
dissolve into flux. In section 4 we analyse the matrix model and
try to recover the Seiberg-Witten curve in the limit where the
superpotential is switched off.
The results in the latter part of section four are somewhat puzzling. Besides the expected
critical point of the glueball superpotential we seem to find some additional critical points. These
can presumably be eliminated by a more careful analysis.

One of the main differences with
the four-dimensional case arises from the possibility of a
Chern-Simons term in five dimensions. The stringy picture gives us a clear intuition for
how to deal with this. Basically there is an integer ambiguity in the glueball superpotential
which is related to the Chern-Simons level. It should be possible to derive this directly from the matrix model,
but we will not attempt this here.
We will also suggest a generalisation of matrix model methods
to theories with more general non-minimal kinetic terms for the gauge fields.

A matrix model for a five dimensional gauge theory with a non-hyperelliptic curve
has been discussed in \HollowoodGR. Recently extensions to six dimensional gauge theories have been discussed in
\HollowoodCV\ and \GanguliKR.

\newsec{A brief tour of geometric engineering}

This section is based on \refs{\KatzFH ,\LawrenceJR,\IntriligatorPQ,\AharonyBH,\LeungTW}.

\subsec{Four-dimensional geometric engineering}

Geometric engineering starts with the observation that IIa string theory on an ALE space gives rise to
a six dimensional gauge theory with sixteen SUSY localised near the `origin' of the ALE. The equation for
an ALE is
\eqn\aa{ uv + x^n =  0 .}
The singularity may be resolved and the middle dimensional homology is then isomorphic to the root lattice
of the algebra of $SU(n)$; it is spanned by $n-1$ 2-spheres with volume
$ |a \cdot \alpha|$ where $a$ is the complexified K\" ahler form on the ALE, and
$\alpha$ runs over the
simple roots of the $SU(n)$ algebra.  The non-abelian gauge bosons are identified with D2 branes wrapped on
various two-spheres in this geometry and their mass is proportional to the volume of the 2-sphere they are wrapped on.

In order to get a four dimensional gauge theory with eight supercharges and no adjoint matter, one fibers the ALE geometry
over a 2-sphere in such a way that the total three-fold is a Calabi-Yau space. The four dimensional
gauge coupling (squared) is then inversely proportional to the area A of the base
$S^2$:
\eqn\aa{ {1\over g_4^2} \sim {A\over g_s \ell_s^2} .}
Further, a massless four-dimensional vector multiplet contains an adjoint-valued
complex scalar  $\Phi$.
It may be expanded in terms of the $SU(n)$ Cartan generators $T_i$ as $\Phi = a_i T^i$, in other
words, the $a_i$ parametrise the Coulomb branch.

String worldsheets wrapped on the 2-spheres act as (point-like) gauge instantons; this may be seen
either from type II on K3/heterotic on $T^4$ duality, or from the fact that in type II
compactified on K3 we have the six-dimensional equation  $d* e^{-2\phi}H = -{\rm Tr} F\wedge
F$.
The instanton number is the number of times the worldsheet configuration is wrapped
on the base ${\bf P}^1$.

The low energy effective action (up to two derivatives) for the massless degrees of freedom
of string theory on this background is determined
by a prepotential ${\cal F}(a_i,A)$
and may be computed by an application of mirror symmetry. After taking a decoupling limit which keeps an interacting
gauge theory but throws away all other modes, one recovers the  gauge theory prepotential  and  action:
\eqn\aa{ 4\pi S =  \int d^4\theta \left({\del{\cal F}\over \del a_i} \bar{a}_i -
\overline{{\del{\cal F}\over \del a_i}}a_i\right)
+ {1\over 2}\int d^2\theta \, {\del^2{\cal F}\over \del a_i \del a_j} W_{i,\alpha} W_j^\alpha + h.c. }
From the gauge theory
perspective, the prepotential consists of three pieces:
${\cal F} =  {\cal F}_{\rm tree}+{\cal F}_{\rm 1-loop}+{\cal F}_{\rm instanton} $, where
\eqn\aa{ \eqalign{
{\cal F}_{\rm tree}&= {1\over 2 } \tau_{0}\, a^t \cdot C \cdot  a \cr
{\cal F}_{\rm 1-loop}&= {i\over 8\pi} \sum_{{\rm roots\
}{\bf \alpha}}({\bf \alpha}\cdot a)^2 \log{({\bf \alpha}\cdot a)^2\over \Lambda^2} -{i\over 8\pi}
\sum_{{\rm flavours}}\sum_{{\rm weights\ }{\bf w}}({\bf w}\cdot a)^2 \log{({\bf w}\cdot a)^2\over \Lambda^2} \cr
{\cal F}_{\rm instanton} &= \sum_{k>0} \Lambda^{Tk}
{\cal F}_k(a_i).}}
We used $\tau_0= {4\pi i \over g_0^2} + {\theta\over 2\pi}$ for the bare gauge coupling, $C$ for the Cartan matrix, and
$T$ for the instanton weight which depends on the gauge group and the matter representations
(it's $2N_c - N_f$ for $SU(N_c)$ with $N_f$ fundamental matter).
 From the stringy point of view
these three pieces receive contributions from classical string theory (here Calabi-Yau intersection numbers),
genus zero worldsheet instantons with zero gauge instanton number with a small piece from the classical intersection
numbers,
and genus zero worldsheet instantons with  gauge instanton number $k$ respectively.

For many gauge theories the relevant Calabi-Yau geometry is actually toric, and the whole procedure may be automated.
Let us discuss the case of  pure $SU(n)$ gauge theory. The linear sigma
model has gauge group $U(1)^{n-1}$ and $n+2$ chiral fields $X_j$. The charge
vectors are given by
\eqn\qmatrix{ \eqalign{
q_b &= (1,1,-2,0,0,0,0,\ldots,0,0,0,0) \cr
q_{f_1} &= (0,0,1,-2,1,0,0,\ldots,0,0,0,0) \cr
q_{f_2} &= (0,0,0,1,-2,1,0, \ldots, 0,0,0,0) \cr
\vdots &= \quad \vdots \cr
q_{f_{n-2}} &= (0,0,0,0,0,0,0,\ldots,1,-2,1,0) \cr
q_{f_{n-1}} &= (0,0,0,0,0,0,0,\ldots,0,1,-2,1) \cr
}}
The sum of the charges in each row is zero; this is the Calabi-Yau condition.
The columns of this matrix can be identified with divisors $D_j$
defined by $X_j=0$. For each row there is a D-term, whose solution set contains
a $T^* {\bf P}^1$ with a minimal size ${\bf P}^1$ with a volume determined by the corresponding FI-term.
Thus to each row we can associate a curve $C_i$,
namely the minimal size ${\bf P}^1$. The first vector corresponds to the base, and the remaining ones
correspond to the exceptional curves in the ALE.
The $(i,j)$th entry of the matrix is the intersection number $C_i \cdot D_j$.
Note that the last $n-1$ vectors
contain the negative of the Cartan matrix for $A_{n-1}$, since it describes the intersections of ${\bf P}^1$'s
in the ALE.

For the purpose of getting a graphical representation of the data and finding the mirror,
one introduces a maximal set of integer vectors $\nu_i$ such that
\eqn\aa{ \sum q\cdot \nu = 0.}
In the present case we can take
\eqn\numatrix{
\nu = \left( \matrix{
1 && -1 && 0 \cr
1 && 1 && 0 \cr
1 && 0 && 0 \cr
1 && 0 && 1 \cr
\vdots && \vdots && \vdots \cr
1 && 0 && n-1 \cr }
\right)
}
The Calabi-Yau condition implies we can always take the first
column to be all 1's. The rows $\nu_i$ are usually represented
graphically by forgetting the first entry and drawing a node
in a plane with position as indicated by the remaining entries. The resulting toric diagram
is sometimes also called a `grid diagram.'

\bigskip
\centerline{\epsfxsize=1.0\hsize\epsfbox{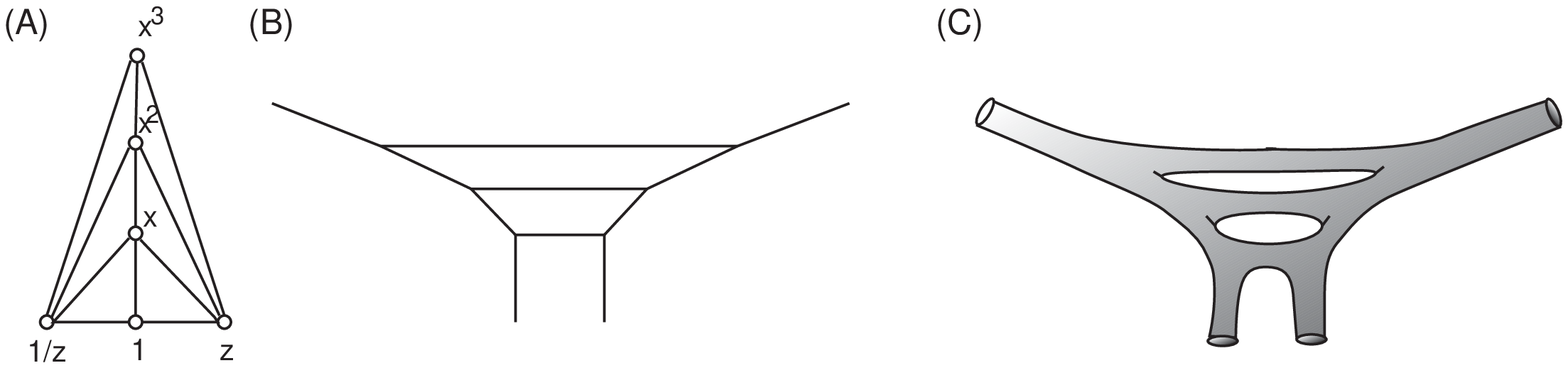}}
\nobreak\noindent{\ninepoint\sl \baselineskip=8pt \fig\toric{}:
{\sl (A): Toric diagram from the data \numatrix\ with $n=3$. This corresponds to
pure $SU(3)$ gauge theory. (B): Dual web diagram. (C): The Seiberg-Witten curve
from thickening the web diagram. }}
\bigskip%

The mirror
geometry is determined by associating a monomial $z^i x^j$ to a node with coordinates $(i,j)$
in the toric diagram. The resulting equation is
\eqn\mirrorcurve{ b_1 z + {b_{-1}\over z}  + \sum_{i=0}^n c_i x^i = 0 .}
One may think of this curve as the `reduction' of the non-compact Calabi-Yau three-fold
\eqn\aa{  b_1 z + {b_{-1}\over z}  + \sum_{i=0}^n c_i x^i =  u\, w .}
Reduction means that the periods of the holomorphic (3,0) form $\Omega = dzdxdu/(zxu)$
over 3-cycles in the Calabi-Yau can be written as periods of the (1,0) form
$\log(x) dz/z$ over 1-cycles of \mirrorcurve.

The independent complex structure parameters in \mirrorcurve\ can be read from
the matrix $q$ \qmatrix:
\eqn\aa{ \eqalign{ z_{f_i} &= {c_{i-1}c_{i+1}\over c_i^2} \cr
z_b &= {b_{-1} b_{1}\over c_0^2 }.}}
If we denote the (complexified) volumes of each of the ${\bf P}^1$'s associated to the charge vectors as
$t_k$, then we have $t_k = -\log(z_k)$.
It is convenient to introduce  flat K\"ahler coordinates $T_k$
which measure physical quantities such as the exact
masses of W-bosons. On the mirror side, the exact masses of BPS particles are computed
by period integrals of $\log(x)dz/z$. These periods
satisfy Picard-Fuchs differential equations. One can then find the $T_k$ explicitly
by solving differential equations and matching the log terms:
\eqn\mirrorreln{ T_k = - \log(z_k) + \sum_{n\geq 0} g_n z_b^{n_b} z_{f_i}^{n_i} .}
The coordinates $T_k$ and $t_k$ differ by
instanton corrections:
\eqn\flatcoord{ T_k = t_k + \sum k_n e^{-n_b t_b} e^{-n_i t_{f_i}} .}
Finally we need to take a limit $\alpha' \to 0$ in which the gauge theory decouples from
other modes but remains interacting:
\eqn\aa{
\exp(-1/g_4^2) \sim \exp(-T_b) \sim \epsilon^{2n} \Lambda^{2n}, \quad
T_{f_i} \sim \epsilon (e_i-e_{i+1}) \quad
{\rm as \ }\epsilon\sim g_s 2\pi \sqrt{\alpha'} \to 0 .
}
In this limit $m_W/\Lambda$ remains finite, $e_i = a\cdot \lambda_i$ where $\lambda_i$ are the fundamental weights
of $SU(n)$, and we take $a_i$ to be the quantum corrected
Coulomb branch parameters (so that $e_i - e_j$ measure exact masses of $W$-bosons).
Then we may write the above equation in terms of physical
parameters as
\eqn\aa{  z + { \epsilon^{2n} \Lambda^{2n}\over z}  +
 \prod_{i=1}^n (x - \exp(\epsilon e_i))  = 0 .}
Putting $x = e^{\epsilon v}$ and taking the limit $\epsilon \to
0$, we obtain
\eqn\aa{z + {\Lambda^{2n}\over z}+ \prod_i (v-e_i) =z + {\Lambda^{2n}\over z}+P_n(v) = 0 .}
Defining $y = 2 z + P_n(v)$, this may be written in hyperelliptic
form:
\eqn\aa{ y^2 = P_n(v)^2 - 4 \Lambda^{2n}.}
Moreover one recovers the expected  Seiberg-Witten differential:
\eqn\aa{ \lambda =  \log(x) {dz\over z} \to  R\,v{dz\over z} .}
This differential can be used to recover the exact prepotential. Namely if we choose a basis of
A-cycles for the curve such that $\int_{A_i} v{dz\over z} = a_i$ and we define the dual B-cycles
by $A_i \cdot B_j = \delta_{ij}$, then we have ${\del {\cal F}\over \del a_j} = \int_{B_j} v{dz\over z}$.

One can introduce fundamental matter by performing a toric blow-up of a point on the
base ${\bf P}^1$. This adds a new ${\bf P}^1$ for which the homology class acts
as a weight of the fundamental representation of $SU(n)$, say the highest weight $\lambda_+$. The other
fundamental weights can be obtained
by adding classes corresponding to the $SU(n)$ roots.
The resulting toric diagram admits several triangulations.
We have drawn two of them below:
\bigskip
\centerline{\epsfxsize=1.0\hsize\epsfbox{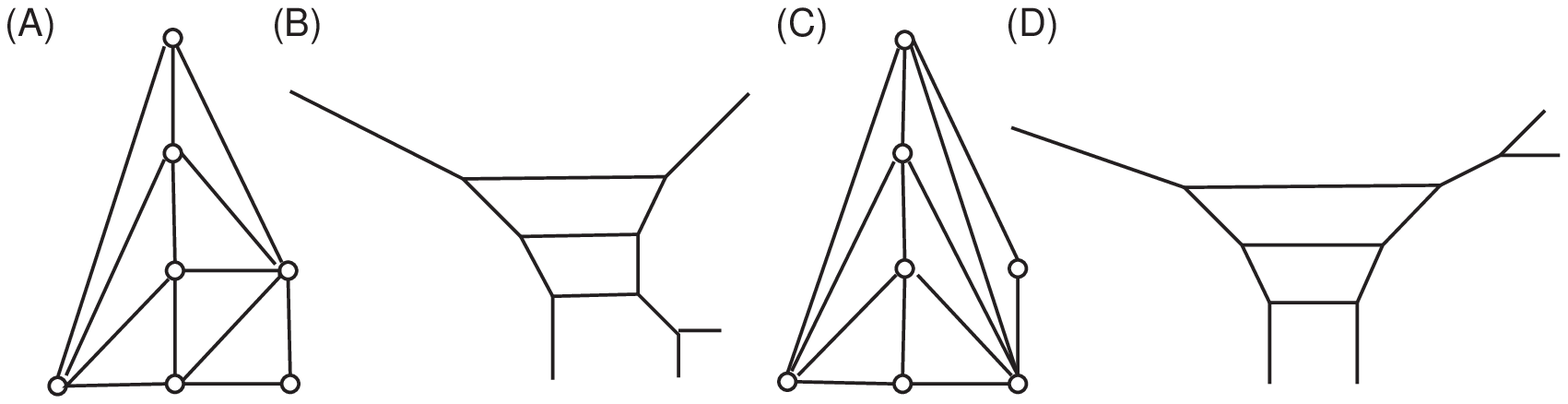}}
\nobreak\noindent{\ninepoint\sl \baselineskip=8pt \fig\transition{}:
{\sl (A): Toric diagram for $SU(3)$ with a fundamental hypermultiplet. (B):
Dual web diagram. (C): A different triangulation of the diagram in (A). (D):
Dual web diagram for (C).}}
\bigskip%
There is an extra parameter corresponding to the mass of the
hypermultiplet $m$. If we take a D2-brane wrapped on the ${\bf P}^1$ with class $\lambda_+$ and
volume $t_{m}$, we will get a particle in four dimensions with mass $ |\lambda_+\cdot a + m|$, so we set
\eqn\aa{ t_{m} = - \log(z_{m}), \quad T_m = R(\lambda_+\cdot a + m) .}
This will yield the Seiberg-Witten curve
\eqn\aa{ z + {\Lambda^{2n-1}(v+m)\over z}  +  \prod_i (v - e_i) = 0 .}

To any given toric diagram one can associate a dual diagram by
 drawing the bi-sectors for
a given triangulation as in \toric B, \transition B and D. This is often
called a (p,q) web diagram. One can
think of the toric Calabi-Yau space as the total space a non-trivial
$T^2$-fibration over ${\bf R}^2$ together with a trivial ${\bf
C}$-fibration. If we identify the ${\bf R}^2$ with the plane of the
dual diagram, then the web tells us where the $T^2$-fibration
degenerates. Namely at each vertex the full $T^2$ shrinks to a
point, and at an edge with tangent direction (p,q) (where $p$ and
$q$ are integers and the vector is primitive) the one-cycle with
class (p,q) in $T^2$ shrinks to a point. We can use dualities to
map this to the corresponding web of (p,q) five-branes in IIb string
theory,
where (p,q) stands for the (R,NS) charge of the five-brane
associated to the edge.
The web picture is often a useful way to think about the
Calabi-Yau.

Strictly the (p,q) fivebranes give rise to five dimensional gauge theories, but
we can get four dimensional theories by compactifying the
worldvolumes of the branes on a circle and shrinking it to zero
size. For finite radius of the circle we can picture this by
thickening the legs of the web diagram, as in \toric C. The
resulting curve is actually the Seiberg-Witten curve.

We may think of the horizontal lines in \toric B as D5-branes. When
they coincide we get an enhanced gauge symmetry;
we denote their vertical coordinate by $e_i \ell_s^2$ (which we can make into a complex number
by adding the Wilson line around the $S^1$ that shrinks to zero size). Then the W-bosons correspond
to strings stretching between the D5-branes and their masses are
proportional to $|e_i - e_{j}|$.

Matter multiplets are introduced by adding semi-infinite
D5-branes, as in \transition B and D, and quantising strings stretched between
the gauge and the matter branes. The vertical coordinate of the matter brane is
related to the real part of the mass of the hypermultiplet (again the
imaginary part comes from a Wilson line).

\subsec{Five-dimensional geometric engineering}

The appearance of $\epsilon = g_s \ell_s = R_{11}$ above is very
suggestive. Indeed it turns out that it is more natural to
consider geometric engineering of five dimensional gauge theories
through M-theory on a Calabi-Yau, and recover the four dimensional theories as a
limit $R_{11} \to 0$ of these gauge theories. This is also hinted
at by the (p,q) five-brane description.

The reasoning is similar. M-theory on an
ALE space again gives rise to non-abelian gauge symmetry (the
non-abelian gauge bosons are now wrapped M2 branes) and fibering
over a ${\bf P}^1$ as before yields gauge theories in five dimensions
with eight supercharges. The action in five dimensions is still determined
by a prepotential which is a function of the vector multiplet moduli $\phi^i$
(these are now real scalars).
The major distinction with the four dimensional case is that the prepotential
has a classical part
and a one-loop contribution, but does not receive
instanton corrections. The triple intersection numbers of the Calabi-Yau, when expressed
in terms of physical quantities,
reproduce exactly the expected prepotential \IntriligatorPQ:
\eqn\cubicpre{ {\cal F} \sim {1\over 2g_0^2}  h_{ij}\phi^i \phi^j + {k_{\rm cl}\over 6} d_{ijk}
\phi^i\phi^j\phi^k +
{1\over 12} \sum_{{\rm roots\ }\bf \alpha} |{\bf \alpha}\cdot \phi|^3
- {1\over 12}\sum_f \sum_{{\rm weights\ }{\bf w}_f } |{\bf w}\cdot
\phi + m_f|^3
.}
Here $h_{ij} = {\rm Tr}(T_iT_j)$ and $d_{ijk} = {1\over 2}{\rm Tr}(T_i(T_jT_k + T_k T_j))$
with $T_i$ the Cartan generators. Moreover $k$ corresponds to the Chern-Simons level in five dimensions:
\eqn\aa{{k_{\rm cl} d_{ijk}\over 24 \pi^2} \int_{R^5} A^i \wedge F^j \wedge F^k .}

When we compactify the five-dimensional theory on a circle, the
prepotential receives additional corrections. We can interpret these corrections
 as one-loop corrections due to a Kaluza-Klein tower of BPS
particles, or alternatively one can interpret the corrections as a one-loop term
in five dimensions plus worldline
instantons of the five dimensional particle. For each particle in five
dimensions with mass $m$ we get a contribution
\eqn\aa{ \tau = {\del^2 {\cal F}\over \del \phi^2} \sim
{1\over 2\pi i} \log\left({1\over R^2 \Lambda_0^2}\sinh^2( Rm/2)\right)  }
with $1/g_4^2 = 2\pi R/g_5^2$. Note that if we put $R\Lambda_0 \sim e^{RM_0/2}$, we recover the usual
five dimensional linear divergence for $1/g_5^2$ as $R\to \infty$.

Since the dilaton lives in a neutral
hypermultiplet and does not talk to the vector multiplets, we can use mirror symmetry
to compute the full prepotential. The crucial difference with the
four dimensional
case is that now when we take the decoupling limit $l_p \to 0$, we do not have to take
a special scaling limit
of the curve. Instead we have
\eqn\aa{ \exp(-T_b) \sim R^{2n} \Lambda^{2n}, \quad T_{f_i} \sim R
M_{W,i} .}
The resulting  Seiberg-Witten curve is
of the form
\eqn\fiveDsw{ z + {\Lambda^{2n}\over z}+  {\rm det}(x-U) = 0 }
with Seiberg-Witten differential
\eqn\aa{ \lambda_{SW} = \log(x){dz\over z} .}
The relation between the coefficients in this equation and the
physical parameters can be determined by solving Picard-Fuchs
equations, with the leading terms at large volume fixed by
\mirrorreln.

Actually \fiveDsw\  is only one of the curves one could get. The gauge and matter
content fixes the toric Calabi-Yau up to a small ambiguity in the
triple intersection numbers \IntriligatorPQ. This ambiguity corresponds to $k_{\rm cl}$
in \cubicpre\ and gives rise to the Chern-Simons term. All these geometries give rise to
the same Seiberg-Witten curve when we take $R\to 0$. The toric diagram we
discussed before (\toric A) corresponds to level $n$, as one can check for instance by computing
the triple intersection numbers of the Calabi-Yau. The other
toric diagrams for the case of $SU(3)$ gauge theory are given
below.
\bigskip
\centerline{\epsfxsize=.7\hsize\epsfbox{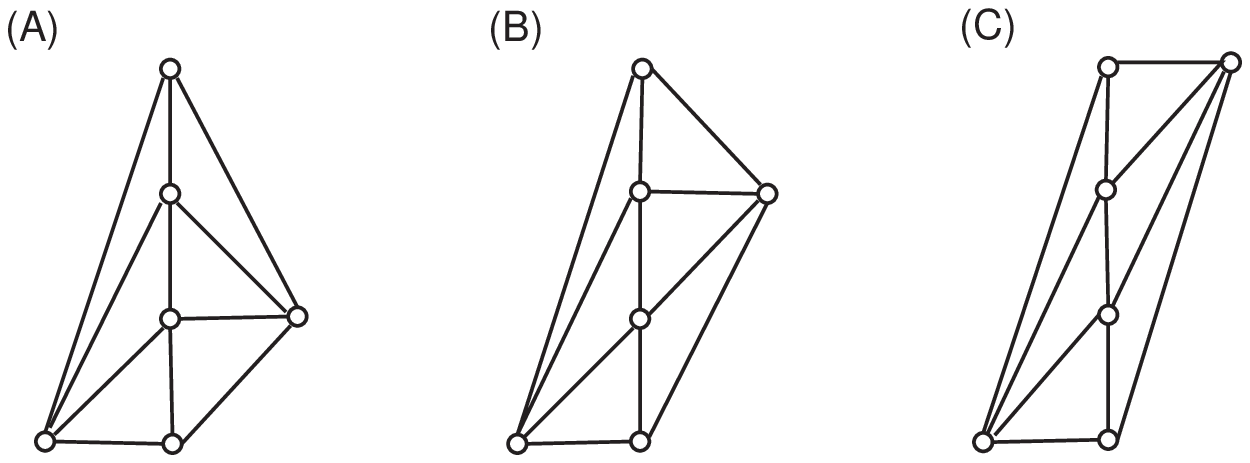}}
\nobreak\noindent{\ninepoint\sl \baselineskip=8pt \fig\csdiagrams{}:
{\sl Toric diagrams for $SU(3)$ gauge theory with (A): level 2, (B): level 1,
and (C): level 0.}}
\bigskip%
There is a simple way to derive the diagrams for different levels from
any given one by adding hypermultiplets. The reason is that
integrating out a hypermultiplet with a large mass $m$ is well known to
give a contribution $\half{{\rm Re}(m)\over
|{\rm Re}(m)|}$ to the Chern-Simons term \WittenQB. Varying the mass between
plus and minus infinity interpolates between successive levels. An
example is given in \transition: as we vary the vertical position
(i.e. the real part of the mass)
of the semi-infinite D5-brane, we cross all of the compact
D5-branes which results in `flops' (changes in the triangulation
of the toric diagram). When we take the semi-infinite D5-brane off
to infinity we can eliminate a node in the toric diagram. This
interpolates between level 3 and level 2 for $SU(3)$.

For general level
$k$ and one hypermultiplet,
the curve may be written  as
\eqn\SWcurve{ z + {(R\Lambda)^{2n-1}e^{Rm/2} x^{n-k_{c}-1/2}(x - e^{-R m})\over z}  = P_n(x)
}
with $P_n(x) = {\rm det} (x - U)$ and Seiberg-Witten differential
$\lambda_{SW} = \log(x) d\log(z)$.
Here $k_{c}$ is the classical CS level
of the five dimensional gauge theory; there is also a one-loop quantum contribution $\half{{\rm Re}(m)\over
|{\rm Re}(m)|}$ from the matter field.
For the case at hand we will put $k_{q} = k_{c} + \half$ which must be an integer
to avoid anomalies. A space-time parity transformation corresponds
to
\eqn\aa{k_{c} \to -k_{c}, \ m \to -m, \ U \to U^{-1}.}
This transformation gives us back the same curve, expressed in the
variable $\tilde{x} = 1/x$ and with a redefined $z$.

Finally for $R|m|$ large compared to $\log(R|\Lambda|)$  we can put $(R\tilde{\Lambda})^{2n} =
(R\Lambda)^{2n-1} e^{\pm Rm/2}$ and take the limit $m\to \pm \infty$ in
\SWcurve. Varying the mass of the hypermultiplet allows us to interpolate between different CS levels
in \SWcurve.

\newsec{Brane/flux duality}

An alternative realisation of five dimensional Yang-Mills is through D-branes. For the case of pure gauge theory
with eight supercharges we can consider type IIa string theory on an $A_1$ ALE space times ${\bf C}^*$ and
wrap D6-branes on the $S^2 \times S^1$.\foot{
Of course this is related to the set-up of the previous section through dualities. A D6 wrapped on
${\bf P}^1$ inside an ALE is essentially T-dual to a D5-brane stretched between to NS5-branes, which gives
the (p,q) five-brane picture. On the other hand as we mentioned in the previous section, this is dual
to geometric engineering. }
The open string description leads to a matrix model, and the planar diagrams of this matrix model
compute an effective superpotential.
The same effective superpotential also arises from the dual closed string description.
We use this setup to discuss some aspects of the  matrix model and effective
superpotential.

For a single D6, the moduli space is
parametrised by the position on the cylinder (the real scalar
in the vector multiplet) and a Wilson line (parametrised by the
dual of the $S^1$). In the following it will frequently be useful to apply a T-duality
on the $S^1$. In particular the modulus of a D6-brane is described by the position of
a D5-brane on the T-dual type IIb
cylinder. For $N$ D6-branes, the moduli space is parametrised by the position of $N$ D5-branes
on the IIb cylinder, the eigenvalues of the complexified Wilson
line\foot{
For convenience we set $R=1$ in this section}
\eqn\aa{
U = P\exp(\int dx_5\, A_5 + \varphi) .}
The Chern-Simons level arises in the following way.
The D-brane action depends on background flux $F^{(2)}_{RR}$ through the
WZ term $\int e^F \wedge\sum C_i$. For a each D6 wrapped on ${\bf P}^1$, we
have
\eqn\aa{ \int_{D6} \! F\wedge F\wedge F\wedge C^{(1)} =
-\int_{R^{3,1}\times S^1}\!\!\!\!\!\!\!\!\!\!\! A\wedge F\wedge F \int_{S^2}dC^{(1)} =
- k \int_{R^{3,1}\times S^1}\!\!\!\!\!\!\!\!\!\!\! A\wedge F\wedge F .}
One may introduce fundamental matter by adding extra D6 branes
wrapped on certain non-compact cycles. If the superpotential is
turned off then the matter D6 brane wraps the $S^1$ on ${\bf C}^*$
and a cycle of the form $x_1=0$ in the ALE
\eqn\aa{ |x_1|^2 + |x_2|^2 - 2 |x_3|^2 = r \ /U(1) .}
This cycle intersects the ${\bf P}^1$ given by $x_3 = 0$ in
precisely one point inside the ALE. The separation on ${\bf C}^*$
with the centre-of-mass of the stack of $N$ D6 branes is related
to the (real part of the) mass of the hypermultiplet, and the
difference between the Wilson lines on the $S^1$ is related to the
imaginary part of the mass.\foot{There are corrections e.g. due to
the warp factor that arises in the presence of space-filling
branes. It is better to think of it as an unobservable bare mass.}
Note that such matter D6 branes act as domain walls for the
background flux $F^{(2)}_{RR}$ on ${\bf P}^1$. Therefore they
affect the CS level depending on the hypermultiplet  mass, and one
can change the value of the background flux by moving matter
D6-branes between $\pm \infty$, hence change the CS level of the
field theory.

Superpotential terms for the pure gauge theory theory living on
the D6 branes may be computed through $U(N)$ Chern-Simons theory on $S^2
\times S^1$:
\eqn\aa{ S_{CS} = {i\kappa \over 4\pi}\int_{S^2\times S^1}
{\rm Tr}\, A\wedge dA + {2\over 3} A\wedge A\wedge A .}
We write the gauge fields as $A_z,A_{\bar{z}}$ and $A_t$, where $t$ is the coordinate along the $S^1$
and $z,\bar{z}$ parametrise the $S^2$. Integrating out gauge degrees of freedom
has the effect of shifting $\kappa \to \kappa + N$. We will require $A_t$ to be independent of $t$, and
identify $g_s = 2\pi i/(\kappa + N)$.
However if there are matter D6 branes or $k$ units of
flux through the ${\bf P}^1$ then the action needs some modification. Let us first consider a single
matter brane with
zero background flux. Strings stretching between the matter and gauge D6-branes
were considered in the topological set-up in \OoguriBV\ and they give rise to a
complex scalar $\phi$. Denoting the coordinate along the $S^1$ by $t$,
the action is modified to
\eqn\aa{ S = S_{CS}+ \int_{S^2 \times S^1} \bar{\phi}(\del_t + A_t + m)\phi
\ \delta^{(2)}_{x_1 = x_3 = 0} .}
The exact normalisation of this term is not relevant here, since
we can rescale $\phi, \bar{\phi}$.

Since we can interpolate between different Chern-Simons levels by
varying the mass of the matter multiplet, let us see how this
modifies the action. Integrating out $\phi$ gives the following term in the action:
\eqn\aa{ -\log {\rm det} \left({\del_t} + A_t +
m\right)= -{\rm Tr} \log \sinh(\pi (A_t + m) ) + {\rm constant}.}
As $m \to \infty$ we kill the $e^{-x/2}$ term in sinh, and as $m\to -\infty$ we
kill the $e^{x/2}$ term. Therefore by starting with pure Chern-Simons
theory and moving branes from minus to plus infinity, we see that
we can account for the background flux by adding a term
\eqn\aa{ -k \int_{S^2 \times S^1} {\rm Tr}\,A_t \ \omega }
to the action,
with $\omega$ a (1,1) form with unit volume on ${\bf P}^1$.\foot{
The equation of motion for $A_t$ then imply
that the gauge field has k units of flux through the $S^2$. This
is similar to the baryon vertex for $k$ quarks where the quarks
are strings stretching from the gauge D6 branes to infinity.}
The periodicity of $A_t$ implies that $k$ is quantised.

We can also turn on a superpotential ${\rm Tr}\, W(U)$ in the gauge theory. One can incorporate this
by turning
on the corresponding term in the Chern-Simons theory, with $U = P \exp \int dt \, A_t$.
The geometric effect of the superpotential is most
easily understood in the T-dual type IIb picture, where it deforms the trivial fibration
of the ALE over the cylinder to a non-trivial one according to the equation
\eqn\superfib{ uw = (y-W'(x))(y+W'(x)).}
Here $x = e^{Rv}$ and the prime denotes differentiation with respect to $v$.
Therefore altogether we consider the partition function for the theory with action
\eqn\topaction{ S = S_{CS} + \int_{S^2\times S^1}dt\, \bar{\phi}(\del_t + A_t + m)\phi\,  \delta^{(2)}_{x_1 = x_3 = 0}
-k \, {\rm Tr}\,A_t\, \omega - {1\over g_s}{\rm Tr}\, W(U)\, \omega }
Note that $A_z,A_{\bar{z}}$ appear only in $S_{CS}$
and appear linearly. Integrating out $A_{\bar{z}}$ gives $F_{tz}=0$ and reduces $S_{CS}$ to
$\int {\rm Tr} A_z \del_{\bar{z}} A_t$. After gauge fixing, this looks
like the topological part of the D5-brane action considered in \DV, and is consistent with the form of
the geometry \superfib\ and the factor of $1/g_s$ in front of the superpotential in \topaction.
Integrating out $A_z$ sets $\del_{\bar{z}} A_t=0$ which implies that
$A_t$ is a constant (we already got rid of the $t$-dependence). Therefore we obtain a
matrix quantum mechanics, and the measure for $A_t$ is that of a unitary matrix model due to the periodic
identifications. The partition function is
\eqn\matrixqm{ Z = {1\over {\rm Vol}(U(N))}\int dU d\phi(t) \, e^{-{1\over g_s}{\rm Tr}\, W(U)
-k {\rm Tr}\,\log U+\int_{S^1}dt\,  \bar{\phi}(\del_t + A_t + m)\phi }}

The geometry \superfib\ contains finite size ${\bf P}^1$'s obtained from blowing up the singular points,
which are in one-to-one correspondence with the critical points of $W$. The D5-branes we get from
T-dualising the D6 branes wrap these  ${\bf P}^1$'s.
In the gauge theory the superpotential Higgses the $U(N)$ to $U(N_1)\times \ldots U(N_n)$ where
$\sum N_i = N$ and $N_i$ is the number of eigenvalues of $U$ that we put at the $i$th critical point
of $W$. In the present set-up
this corresponds to wrapping $N_i$ D5-branes on the ${\bf P}^1$ we get from blowing up the $i$th singular point
of \superfib.

We can alternatively describe this system as a closed string theory on a Calabi-Yau which is obtained by excising
each ${\bf P}^1$ and glueing in an $S^3$. In the dual geometry the D-branes are replaced by fluxes. From the perspective
of the field theory this is a large $N$ dual description in terms of glueball fields
$S_i \sim {\rm Tr_{U(N_i)}}W_\alpha W^\alpha$. The scalar component of the superfield $S_i$ is related to the size
of the $i$th $S^3$ in the dual geometry
and therefore corresponds to a complex structure parameter of the dual geometry.

In  the closed string set-up the Calabi-Yau
three-fold is given by\foot{More accurately, the space-time metric
also includes a warp factor that takes into account the
backreaction of the non-zero fluxes.}
\eqn\aa{ uw - y^2 + W'(x)^2 + q(x) =0 }
where $q(x)$ corresponds to a (log-)normalisable complex structure deformation of
\superfib\ which gives finite volume to the $S^3$'s.
The complex structure parameters are governed by an effective superpotential\foot{
Physically to find the effective superpotential we should expand around the space-time
solution
which includes the warp factor and integrate out all unwanted fields.
 The standard claim in the literature (eg. \VafaWI) is that the
effective superpotential is defined on the moduli space of the original Calabi-Yau before
any fluxes were turned on. This requires a remarkable cancellation of the warp factor. I would
like to thank V.~Zhukov for a conversation on this long ago.}:
\eqn\aa{ W_{\rm eff} \sim \int_{{\rm CY}_3} H \wedge \Omega }
where $H = H_{RR} + \tau H_{NS}$ and $\Omega$ is the holomorphic 3-form on the Calabi-Yau on the IIb side.
One can effectively reduce the superpotential  to\foot{
The correct normalisation should be $W_{\rm eff} = -{1\over 2} \int_{\Sigma} h \wedge \lambda$ so that in the $g_s N \to 0$ limit
(in which $q(x) \to 0$) we get the expected superpotential $\sum N_i W(x_i) + C$,
with $x_i$ the critical points of $W$ and $C$ depends on the cut-offs $\Lambda_\pm$ but not
on the $x_i$.
Nevertheless we will not explicitly include the $-{1\over 2}$ below.}
\eqn\fluxpot{ W_{\rm eff} \sim \int_{\Sigma} h \wedge \lambda }
where $\lambda = \sqrt{W'(x)^2 + q(x)}\, dx/x$, and $\Sigma$ is the curve defined by
\eqn\aa{ y^2 = W'(x)^2 + q(x) .}
A representation of this curve is given \contours.

Expanding $W_{\rm eff}$ in terms of the periods of $h$ and $\lambda$ gives us an effective superpotential for
the glueball fields $S_i$. It is not entirely clear we can interpret it as an effective superpotential in the
conventional sense since we are in a massive vacuum and the $S_i$ may not be the lightest degrees of freedom. However
it clearly has non-trivial information because we use it to compute exact tensions of BPS domain walls. Moreover
after minimising, solving for the VEV's of the $S_i$ and substituting back,
we can interpret the resulting superpotential as a generating function for the quantum expectation values
$\vev{{\rm Tr}( U^m)}$.\foot{
Actually the effective superpotential refers to a particular UV completion of the field theory. See
\refs{\AganagicXQ,\KennawayJT} for a discussion.}
The reduction of the open string theory to the matrix quantum mechanics \matrixqm\ implies that the effective
superpotential can be computed from the matrix quantum mechanics.

In order to expand \fluxpot, we would like to know
the periods of $h$. This works similar to the four dimensional case,
with a twist because the Riemann surface has two extra punctures. The integral of $h$ around each cut
should give the number of eigenvalues we put at the corresponding critical point of $W$. What about the other periods?
\bigskip
\centerline{\epsfxsize=.7\hsize\epsfbox{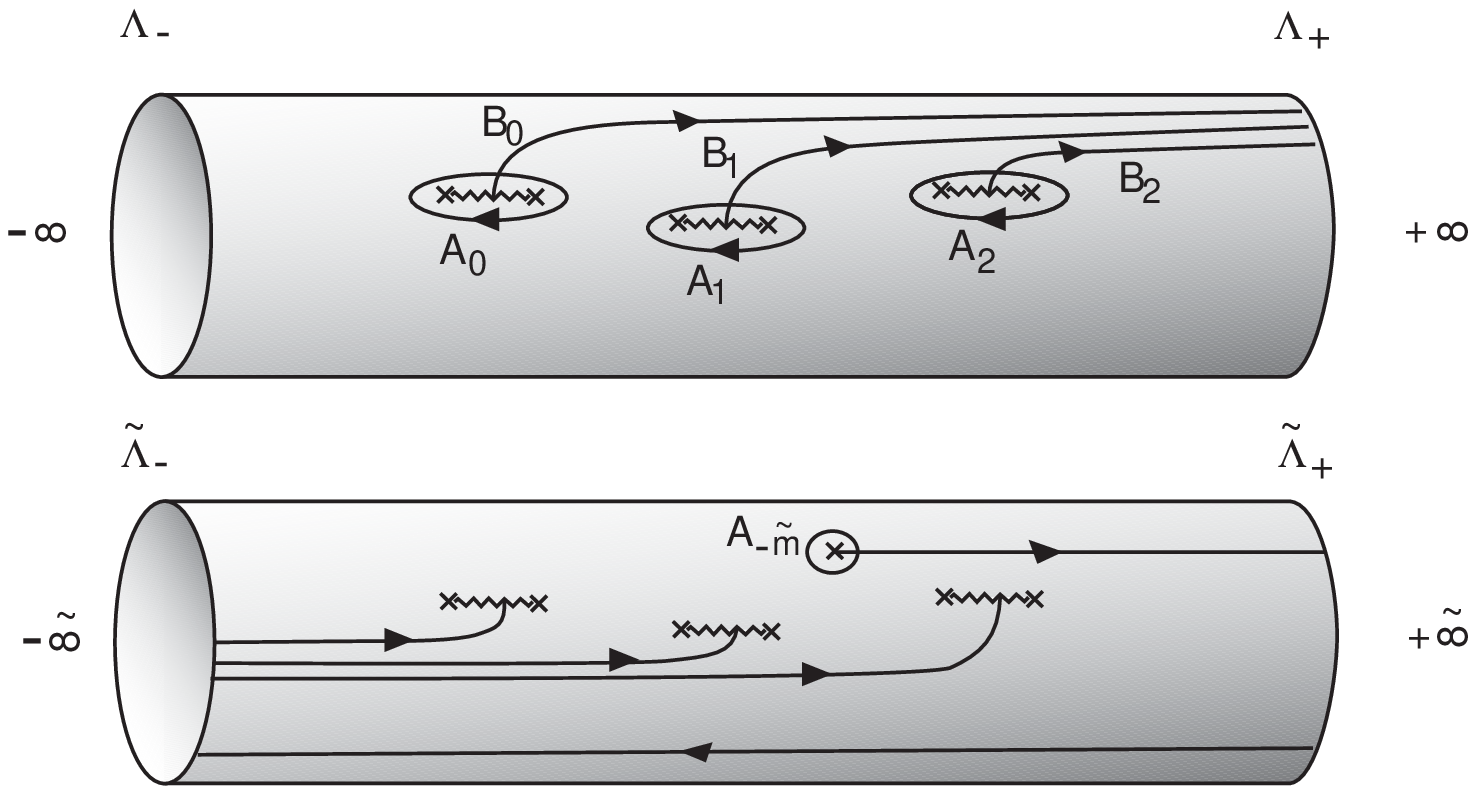}}
\nobreak\noindent{\ninepoint\sl \baselineskip=8pt \fig\contours{}:
{\sl The matrix model curve $y^2 = W'(x)^2 + q(x)$ as the double cover of a cylinder parametrised
by  $x = e^{Rv}$. The B-cycles
essentially follow flux lines which are discussed in the text.  The geometry of the curve looks very
similar to a Seiberg-Witten curve as in \toric C turned $90^0$, but they are unrelated at this point. }}
\bigskip%

After T-duality, the background flux $F^{(2)} \sim k \, \omega_{S^2}^{(2)}$ gives rise to
a background 3-form flux. Upon reduction to $\Sigma$, this gives rise to a background flux for $h$.
Similarly any D6 brane is a magnetic source for 2-form flux; after
T-duality the matter D6 branes gives rise to additional sources for 3-form flux, which in turn gives
pointlike sources for
$h$ after reduction.
Therefore the flux $h$ has a constant term $k$ related to the CS level
and sources at the position of the matter D5-branes.

The background flux and the matter sources appear on the `x-plane' (cylinder); the algebraic curve
obtained from reduction of the Calabi-Yau is a double cover of that. So we should decide on which sheet
to put the matter sources and the background flux for $h$. Let us first understand how it works
for zero background flux and no matter.

For $N$ gauge D5-branes we get $N$ charged particles (or matrix model eigenvalues)
on the cylinder. The fluxes through $\pm \infty$ on each sheet
must therefore add up to $\pm N$. On the upper sheet we have to make a choice to distribute $f$ units of flux
at $-\infty$ and $N-f$ units of flux at $+\infty$. No physical quantities will depend on this choice, but
once we make it all other fluxes will be fixed. We will choose $f=0$. Then the gauge D5-branes are effectively
attracted to a particle at $+\infty$ with charge $N$. In order to balance this force, in the absence
of background flux or other sources, on the lower sheet we must then put the $-N$ units of flux entirely
at $-\tilde{\infty}$. Not doing this would lead to an additional force term in the matrix model action.

Now we introduce matter branes. Recall that classically we have
$\vev{{\rm Tr}(U^r)} = \oint x^r h(x)dx$ where $h$ has poles at positions corresponding to
the classical positions of the $N$ gauge D5 branes, and is independent of the position of
additional matter branes (if the
hypers don't get a VEV). Therefore $h$ should get contributions from
matter only on the `quantum' sheet that is not visible classically, and we take this to be the lower sheet.\foot{
If some of the hypers get a VEV, the poles do appear on the upper sheet \CachazoYC.}
Then the background flux $k$ will also only appear on this sheet, because we can create it by moving matter branes around
on the second sheet. This fixes all
the A-periods of $h$.

Finally let us discuss the B-periods of $h$. As in the
four-dimensional case we will take the compact B-periods to be
zero (although this can be generalised slightly as in \CachazoYC). The
non-compact periods are divergent however so we need to introduce
cut-offs to define the integrals. In principle there are four
asymptotic infinities and we need to introduce a cut-off for each
of them. This might seem very strange from the gauge theory point
of view where we need only one cut-off. However it is by now a
familiar story that the gauge theories we are analysing are not
renormalisable and the matrix model seems to refer to a particular
UV completion \refs{\AganagicXQ,\KennawayJT}.
Recently it was understood in the four dimensional
case \KennawayJT\ that introducing cut-offs for the divergent integrals
in the effective superpotential is
indistinguishable from
introducing sufficiently many matter branes on each sheet at the position of the cut-offs,
so that the total matter content is $\tilde{N}_f = 2 N_c$.
In other words, one can think of the cut-offs as the masses of additional
matter multiplets. It appears that this
explanation is valid in the five-dimensional case also. Indeed we have already observed that we effectively
have $N_c$ charges at infinity on the upper sheet and $N_c - N_f$ charges at infinity on the lower sheet (for any
Chern-Simons level).
We can identify these charges with additional matter branes, thus embedding the original theory into a theory
with total number of flavours $\tilde{N}_f = 2 N_c$. Interestingly, the underlying ${\cal N}=2$ theories are also
precisely the theories that can be lifted to six dimensions, which suggests that should be the natural starting point.

At any rate, we will use a cut-off $\Lambda_+,\tilde{\Lambda}_+$ for $+\infty, +\tilde{\infty}$ and
$\Lambda_-,\tilde{\Lambda}_-$ for $-\infty, -\tilde{\infty}$. Due to the special
form of the A-periods of $\lambda$ that we will derive later, it
turns out that the glueball superpotential depends only on the
combination
\eqn\aa{ \int_{-\tilde{\infty}}^{-\infty} h +
\int_{+\tilde{\infty}}^{+\infty} h  .}
To get the correct running of the gauge
coupling it appears that we need
\eqn\aa{ {1\over 2}\int_{-\tilde{\infty}}^{-\infty} h + {1\over 2}
\int_{+\tilde{\infty}}^{+\infty} h =  \tau }
which is consistent with the $R\to 0$ limit.

\newsec{The matrix model.}

In the last section we have indicated from the string set-up that we expect to be able to
compute F-terms of the five-dimensional gauge theory in terms of a `dimensionally reduced' theory.
One can also argue directly from the field theory
perspective that such a reduction occurs \DijkgraafXK.

We are interested in the $F$-terms, which according to \ArkaniHamedTB\
can be written is 4D superspace as
\eqn\aa{
\int d^5x \int d^2\theta \,{\rm Tr}\ (\tau + k\Phi) W_\alpha W^\alpha
+ H^c(\del_t +\Phi + m) H + {\rm Tr}\, W(U) + {\rm h.c.}
}
Now the logic of \DijkgraafXK\ runs as follows: if we think of the coordinate $t$ as a label,
then we can think of the 5D theory as a 4D theory with an infinite number of fields.
Then applying what we know about computing F-terms in a 4D theory, we expect that we need to consider a bosonic
quantum mechanics with action given by%
\eqn\aa{
{\cal S} = -{1\over R g_s} \int dt \,\left[ H^c (\del_t + \Phi + m) H  + {\rm Tr}\,W(U)\right] .}
Here $U = e^{R \Phi}$
with $\Phi \sim \Phi + 2\pi i /R$. As in the topological theory, we take $\Phi$ purely imaginary
so that $U$ is unitary, and $H^c = \bar{H}$. Since $H^c,H$ appear quadratically,
we can immediately reduce this to a matrix model.
Rescaling  $H^c(t),H(t)$ and integrating them out, we get the partition function
\eqn\aa{
Z \sim {1\over {\rm Vol}( U(N))} \int dU\, {e^{-{1\over g_s} {\rm Tr}
W(U)} \over {\rm det} (\del_t + \Phi + m)} .
}
Actually we only expect this for $k=0$. For $k\not= 0$ there may be a modification.
Just as in the previous section we can find the full answer by adding some hypermultiplets and taking
their masses from minus to plus infinity. In general we get
\eqn\aa{
Z \sim {1\over {\rm Vol}(U(N))} \int dU\,{\rm det}(U)^{-k} \, {e^{-{1\over g_s} {\rm Tr}
W(U)} \over {\rm det} (\del_t + \Phi + m)} .
}
In other words, when we take the $d^2\theta$ terms in the field theory action
as the action for the matrix model,
we should be careful to include a contribution $k {\rm Tr}(\Phi)$
for the Chern-Simons term.  This is not unreasonable\foot{ I would like to thank C. Vafa
for this remark} given that
the field theory resolvent ${\rm Tr}\, W_\alpha W^\alpha /(z-\Phi)$
(plus images if $\Phi$ is periodic) is replaced in the matrix
model by ${\rm Tr}\, 1/(z-\Phi)$.

It is tempting to state this in general: non-minimal kinetic terms of the form
\eqn\aa{ \int dx d^2\theta\,  {\rm Tr}( f(\Phi) W_\alpha W^\alpha)}
should lead to a term ${\rm Tr}(f(\Phi))$ in the matrix model action (without a factor
of $1/g_s$). This can perhaps be shown
in a similar way by integrating out matter with appropriate mass terms.

Now we proceed to apply some standard techniques.

\subsec{Deriving the matrix model curve.}

Diagonalising $U = {\rm diag}(e^{R \phi_1}, \ldots, e^{R
\phi_N})$ and using the formula
\eqn\aa{
{\rm det}(\del_t + x) \sim {\rm det}{\sinh ({1\over 2}R x) \over
{1\over 2}R} }
we obtain
\eqn\aa{\eqalign{
Z &\sim  {1\over {\rm Vol}} \int dU\, {e^{-{1\over g_s} {\rm Tr}
W(U) - k {\rm Tr} \, R\Phi} \over {\rm det} \sinh \hb ( \Phi +  m)} \cr
&\sim {1\over {\rm Vol}} \int \prod d\phi_i \,
{\prod_{i\not = j}\sinh \hb(\phi_i - \phi_j) \over \prod_i \sinh \hb ( \phi_i +  m)}
 e^{-{1\over g_s} \sum_i W(R\phi_i) - k \sum_i R\phi_i} \cr
&\sim
{1\over {\rm Vol}} \int \prod d\phi_i \,
e^{-{1\over g_s} \sum_i W(R\phi_i)
+ \sum_{i\not = j}\log \sinh \hb(\phi_i - \phi_j) - \sum_i \log \sinh \hb (m+\phi_i) -k R \phi_i}
.}}
Notice that the Chern-Simons level manifests itself as an additional force term in the action, as expected
from our discussion in section 3.
The equations of motion are given by
\eqn\eom{
-{1\over g_s} W'(R \phi_i) + \sum_{j,j\not= i}\coth
\hb(\phi_i-\phi_j) - {1\over 2}\coth \hb(m+\phi_i)-k  = 0 .}
Introducing the resolvent
\eqn\aa{
\omega(v) = -{1\over N}\sum_i \coth \hb(v-\phi_i) }
we can bring expression \eom\ into the form
\eqn\loopeqn{\eqalign{ 0 &= -{1\over g_s N} f(v) + {1\over g_s
N}W'(R v) \omega(v) + {1\over 2}[\omega^2(v) - {2\over R
N} \omega'(v) -1] \cr & \qquad -{1\over 2 N^2} \sum_i \coth \hb(m +
\phi_i) \coth \hb (v-\phi_i) + {k\over N}\omega(v) }}
where the error term is given by
\eqn\aa{
f(v) = {1\over N}\sum_i\left(W'(R \phi_i) - W'(R v)\right)
\coth\hb(v-\phi_i) .}
In the large $N$ limit, with $S = g_s N$ fixed, this reduces to
\eqn\aa{ -{1\over g_s N} f(v) + {1\over g_s N}W'(R v)
\omega(v) + {1\over 2}[\omega^2(v)-1] = 0.}
Defining
\eqn\aa{
y = S \omega(v) + W'(R v) }
this can equivalently be written as
\eqn\mmcurve{
y^2 = (W'(R v))^2 + 2 S f(v) +S^2 = (W'(R v))^2 + q(v).}
As mentioned before $q(v)$ must corresponds to a normalisable
complex structure deformation. This means that, if $b$ is a
parameter in $q(v)$, then the integral of holomorphic form $\del_b
\lambda$ over a contour that runs to infinity must be convergent
or at most logarithmically divergent. One of the differences of
the five dimensional case compared to the four dimensional case
is the existence of two asymptotic infinities (on each sheet).
As a result of this we will see that there are two logarithmically
divergent (1,0) forms as opposed to only one in the four
dimensional case. However we will see only a particular
combination of the two arises from the matrix model. Thus it
appears that not all allowed complex structure deformations
actually occur.

\subsec{Relations between the coefficients.}

A simple counting shows that there must be relations between the $b_i$.
We consider two cases in turn\foot{
In the first case $W$ is a holomorphic function
with a pole only at infinity. Therefore picking a `contour' which encloses no poles in the
evaluation of the partition function,
such as integration over unitary matrices, gives identically zero.
This should not be relevant however, since for field theory purposes we are not interested
in defining the matrix model non-perturbatively. I would like to
thank C. Roemelsberger for this remark.}:
$W(U)$ has only positive powers of $U$, or $W(U)$ contains both positive and
negative powers of $U$.

\noindent
${\underline {\rm case \ i}.}$ $W(U) = \sum_{i=0}^{n+1} g_i U^i.$

In this case $W$ has $n$ critical points at finite distance and one at $-\infty$ (from solving
$W'(v) \sim x\del W/\del x = 0$). It is easy to see that in this case the quantum correction is of the form
\eqn\aa{q(x)=\sum_{i=0}^{n+1} b_i x^i .}
Moreover one can see that the constant term in $f(x)$ is zero because $\vev{ U\del W/\del U}=0$, and the leading term
in $f(x)$ is $(n+1)g_{n+1}x^{n+1}$. Therefore there is a relation between $b_0$ and $b_{n+1}$:
\eqn\casei{ {b_{n+1}\over 2(n+1)g_{n+1}} = \sqrt{b_0} = S .}
Moreover this has the following implication for the periods of $\lambda_{mm} = y\, dx/x$:
\eqn\aa{ {1\over 4\pi i}\oint_{v=\infty}\!  \lambda = {1\over 4\pi i} \oint_{v=- \infty}\!  \lambda = S/2 .}
The contours around $\pm \infty$ are taken counterclockwise.
Taking account of \casei,
this implies that the quantum correction has
$n+1$ adjustable parameters. This is the right number: we have a total of $n+1$ critical points
which get smeared out into cuts. Here we include the cut that degenerates to a double root
at $-\infty$ as the quantum correction is turned off.
Let us denote by $A_i$ a contour which encircles the $i$th cut (we associate $A_0$
with the root at $-\infty$). Then we have $n+1$ independent
variables $S_i$ which are related to the $b_i$ through
\eqn\aa{{1\over 4\pi i} \oint_{A_i} \lambda = S_i \quad i = 0, \ldots, n.}

\noindent
${\underline {\rm case \ ii}.}$ $W(U) = \sum_{i=-1}^{n} g_i U^i.$

(we could allow for more powers of $U^{-1}$, but this doesn't add
anything new). In this case the superpotential has $n+1$ critical
points. However $W$ has only $n$ parameters to adjust the
positions of the roots (the other is an overall scalar
multiplying $W$). Thus the position of the last critical point is
fixed in terms of the previous $n$.

The matrix model again implies a relation between the coefficients
of the quantum correction. In this case it is easy to determine
the highest and lowest terms in $f(x)$. We find that
\eqn\caseii{ {b_n\over n g_n} = {b_{-1}\over g_{-1}} = 2S }
so that again we have
\eqn\aa{ {1\over 4\pi i}\oint_{v=\infty}\!  \lambda = {1\over 4\pi i}\oint_{v=- \infty}\!  \lambda = S/2 .}
This gives the right
number of parameters: there are $n+1$ independent $b_i$ after
taking into account \caseii, and there are $n+1$ variables $S_i$
related to the $b_i$ through
\eqn\aa{ {1\over 4\pi i} \oint_{A_i} \lambda = S_i \quad i = 0, \ldots, n.}
Here we take the contours $A_i$, $i = 1, \ldots, n$ to encircle $n$ independent
critical points and the contour $A_0$ to encircle the remaining critical point.

\subsec{Effective superpotential }

The effective superpotential is given by the following formula
\eqn\aa{ W_{\rm eff} = \int h \wedge \lambda}
where $\lambda = y\, dx/x$. We have discussed previously the
correspondence between the field theory data and the periods of
$h$ using the string set-up. One should also be able to derive this directly from the matrix model
as in \NaculichHR, which was made more precise in appendix B of \CachazoYC.

The A-periods of $h$ were as follows:
\eqn\hfluxes{ \matrix{
\oint_{A_i} h = N_i && \oint_{A_0}h = N_0 &&
\oint_{-\infty}h = 0 &&
\oint_{\infty}h = N \cr
\oint_{-\tilde{\infty}}h = -(N-k_{q}) &&
\oint_{\tilde{\infty}}h = -(k_{q}-1) && \oint_{A_{-\tilde{m}}}h = -1 }.}
We use $\pm \infty$ for the two ends of the cylinder, and a tilde for points on the second sheet
(which cannot be seen classically). Also $N = \sum N_i$,
and $A_{-\tilde{m}}$ is a contour
around the point $v = -m$ on the lower sheet.

We also need to specify the periods for the B-cycles. We take
$\int_{C_i} h = 0$ for the compact cycles (where $C_i = B_i - B_0$)
and\foot{
The precise contour we take is not relevant since $\int_{C_i} h =
0$. In cases where it is ambiguous, contours written as $a\to b$ mean
the following: a contour between two points on the same sheet
is chosen to lie on that sheet, and a contour between two points
on different sheets is chosen to cross the $A_0$ cycle but no
other cycle $A_i$ for $i>0$.}
\eqn\hnoncompact{ \half \int_{-\tilde{\infty}}^{ -\infty} h +  \half\int_{+\tilde{\infty}}^{
+\infty} h = \tau.}
We
believe this is the correct prescription by comparing with the four dimensional limit of
the five dimensional gauge theory.

Now we expand the superpotential and obtain
\eqn\aa{\eqalign{
W_{\rm eff} & = \sum_{i=1}^n N_i \int_{C_i} \lambda
+ N\int_{-\tilde{\infty}}^{+\infty} \lambda +
k \int_{+\tilde{\infty}}^{ -\tilde{\infty}} \lambda +
\int_{ -\tilde{m}}^{+\tilde{\infty}}\lambda + 4\pi i\tau\sum_{i=0}^n
S_i \cr
&= \sum_{i=0}^n N_i \int_{B_i} \lambda
 + k \int_{+\tilde{\infty}}^{ -\tilde{\infty}} \lambda +
\int_{ -\tilde{m}}^{+\tilde{\infty}}\lambda + 4\pi i\tau\sum_{i=0}^n
S_i . \cr}}

As mentioned there is in principle some more flexibility in the periods of $h$.
We have chosen to distribute the $N$ units of flux over the two infinities
on the upper sheet in a particular way; we can shift the fluxes to
$\oint_{-\infty}h = f$ and
$\oint_{\infty}h = N -f$ provided we also shift the flux at $-\tilde{\infty}$ and $+\tilde{\infty}$ by $-f$ and
$+f$ respectively. This should have no effect on the effective superpotential when written in
terms of physical parameters only. Finally we can move the matter brane through a cut onto the classical
sheet and take it to the ends of the cylinder, changing the fluxes
at $\pm \infty$. This corresponds to a Higgsed phase of the field
theory \CachazoZK.

\subsec{Minimisation of the superpotential, and factorisation of
the curve.}

In the remainder we would like to apply the strategy employed in
\CachazoPR\ to extract the Seiberg-Witten curve. The idea is to
pick an arbitrary point on the Coulomb branch, i.e. choose
$N$ points on the cylinder, and consider a superpotential $W$ for which
these $N$ points are all critical points.\foot{
Actually we will see that in the five dimensional case the relation between
the point on the Coulomb branch and the critical points of $W$ is a little more complicated.}
Then in the matrix model
we put one eigenvalue at each critical point. After minimising the
effective superpotential, we can scale the superpotential to zero.
The $U(1)^N$ coupling constants are unchanged under the rescaling
of the superpotential and so agree with the coupling constants
of the ${\cal N}=2$ theory obtained from the Seiberg-Witten curve.

In the four dimensional case it was possible to choose $W'(x) =
{\rm det}(x-\Phi)$ (for pure gauge theory), so that the matrix model curve
reduces to the Seiberg-Witten curve up to a rescaling.
For the
five-dimensional case however, for the superpotentials considered
above we have seen that we cannot choose all the critical points
independently.

 We will now try to recover
the Seiberg-Witten curve
by choosing a classical superpotential $W$ which has $N$ independent critical points
and an additional dependent critical point (corresponding to the
$A_0$ cycle for cases i and ii above).

Thus we take $N_0 = 0,N_{i>0} = 1$ and minimise with respect to the $S_i$.
We have noted before that there are $n+1$ independent
variables $S_i$ and $n+1$ independent coefficients $b_i$ in
$q(x) = \sum x^i b_i$. So provided the Jacobian for the change of variables is non-singular,
we can equivalently minimise with respect to the $b_i$.
We expect that when we solve and substitute back, the cut where we did not put any eigenvalues will
degenerate to a double point, so that the matrix model curve factorises as
\eqn\aa{
y^2 = H(x)^2 F(x) .}
The details for cases i and ii introduced above will diverge due to the fact that the
relation between the $b_i$'s is slightly different. We will treat case ii
here.

For case ii, $\del_{b_i} \lambda$ for $i=0, \ldots , N-1$ span the space of holomorphic 1-forms on the matrix model
curve, which has genus $N$. For $i=-1,N$ we find that $\del_{b_i} \lambda$ is logarithmically divergent at one of
the ends of the cylinder, but we can
not consider these two independently because of the relation between $b_{-1}$ and $b_N$ \caseii.

Differentiating $W_{\rm eff}$ with respect to $b_j$ for
$j=0,\ldots n-1$ we find that
\eqn\aa{ \sum_{i=1}^n N_i \int_{C_i} \del_{b_j}\lambda = 0}
and
\eqn\aa{N\int_{-\tilde{\infty}}^{+\infty} \del_{b_j}\lambda +
k_q \int_{+\tilde{\infty}}^{ -\tilde{\infty}} \del_{b_j}\lambda +
\int_{ -\tilde{m}}^{+\tilde{\infty}}\del_{b_j}\lambda =0 .}
Invoking Abel's theorem, this implies there must exist a meromorphic function
on the matrix model curve with a pole of order $N$ at $\infty$, a zero of order $k_q-1$ at $+\tilde{\infty}$
a zero of order $N-k_q$ at $-\tilde{\infty}$
and a zero of order 1 at $-\tilde{m}$, and regular elsewhere. Such a function is of the form
\eqn\zdef{ z\sim P_N(x) + \sqrt{P_N(x) - \gamma x^{N-k_q}(x-e^{-R m})} .}
with $P_N(x) = \sum_i p_i x^i$ a polynomial in $x$ of degree $N$. We take $p_N = 1$ and $p_0 \not = 0$.
Because of the square root, such a function can only exist on a curve of genus $N-1$.
Then the matrix model curve must factorise as
\eqn\factorcurve{ y^2 = ( x\, \del_x W)^2 + q(x) = g^2(x^{-1}-a)^2 (P_N^2(x)
- \gamma x^{N-k_q}(x-e^{-Rm})) .}
The parameter $a$ can be used to ensure that the condition \caseii\ for $b_{-1}$ and $b_N$
is satisfied. The constant $g$ is an overall scale in the superpotential.
We can take $g\to 0$ to turn off the superpotential without changing
the positions of the critical points. The last available parameter $\gamma$ is  fixed by solving
\eqn\delS{\del_S W_{\rm eff} \equiv 2 (N g_N\del_{b_{N}}+g_{-1}\del_{b_{-1}})W_{\rm eff} = 0 .}

Clearly the last factor in \factorcurve\ looks a lot like the expected Seiberg-Witten curve. It needs
to be shown that for a given $P_N(x)$ one can construct a superpotential $W(x)$ with the property that after minimisation
of the effective superpotential $W_{\rm eff}(S)$ the matrix model curve factorises as in \factorcurve\
with the expected polynomial $P_N(x)$. To simplify the algebra a bit, let us assume there is no
matter multiplet. Then we have
\eqn\factorcurve{ y^2 = (x\,{\del_x}W )^2 + q(x) = g^2(x^{-1}-a)^2 (P_N^2(x)
- \gamma x^{N-k}) .}
As an Ansatz for the relation between $W'$ and $P_N$ we write
\eqn\aa{   x\, {\del_x}W  = g(x^{-1}-a)P_N(x) + g c + g d x^{-1} .}
We can get an equation for the constant $c$  by requiring that the right-hand-side
has no constant term
(otherwise we would get a $\log(x)$ term in $W$, which does not respect the periodic identifications imposed
by gauge invariance). So
\eqn\aa{ c = -  p_1 +  a p_0.}
The constant $d$ is fixed by requiring that there is no $x^{-2}$ term in
$q(x)$.
Expanding $x\, \del_x W$ in \factorcurve\ we find that
\eqn\aa{ q(x) = -2g^2(x^{-1}-a)(c+dx^{-1})P_N(x)-g^2(c+dx^{-1})^2
-g^2(x^{-1}-a)^2\gamma x^{N-k} .}
One can check it is always possible to satisfy \caseii\ for any of the allowed values $k = 0, \ldots, N$.
So for every allowed $P_N(x)$ there exists a superpotential $W(x)$
for which \factorcurve\ is realised as a critical point of the effective
superpotential $W_{\rm eff}(S)$.

For $k=0$ we find $d=0, c=\gamma p_1/(\gamma+4p_0)$. For `generic' $k$ ($k=1,\ldots, N-1$) we seem to find two branches.
If $d=0$ then also $c=0$, which looks like the vacuum we are after. But if $d=-2p_0$ then $c$ can be anything. The latter branch seems puzzling because
we have used $N-1$ equations but still have two undetermined variables ($c$ and $\gamma$), so the
glueball superpotential appears to have a flat direction. Moreover on this branch we fail to have the property
that $c,d\to 0$ as $\gamma \to 0$ so that the roots of $P_N$ do not `classically' coincide with critical points of
$W$. We suspect this branch can be eliminated by a more careful analysis.

For $k=N-1$ we find $d=0$ and $c=\gamma/(4p_0)$. Finally for $k=N$ we find two solutions,
depending on the sign in $d = -p_0 \pm \sqrt{p_0^2 + \gamma}$. Requiring $c,d\to 0$ as $\gamma \to 0$ singles
out $d= -p_0 + \sqrt{p_0^2 + \gamma}$ as the expected solution, and the other solution should presumably
be eliminated.

Notice that for several values of $k$ we found that the roots of $P_N(x)$ are slightly
displaced with respect to the critical points
of $W(x)$. This should presumably be interpreted as arising from instanton corrections to the classical
expectation values $\vev{{\rm Tr}(U_{\rm cl}^m)}$.

Thus it appears we have recovered the Seiberg-Witten curve from
the matrix model curve \factorcurve. However this is meaningless if we don't
recover the correct Seiberg-Witten differential, which amongst
others gives the $U(1)$ coupling constants and the identification of the coefficients of $P_N$
with Coulomb branch parameters in the ${\cal N}=2$ theory.

\subsec{Seiberg-Witten differential}

Recall that in the string theory embedding the D-branes are replaced in the large $N$ limit
by fluxes. Therefore in the effective theory
the gauge theory density of eigenvalues of $\Phi$ is given by $h$, which can be used to compute quantum expectation values. On the other hand the Seiberg-Witten
differential does the same thing, so  $\lambda_{SW} = \log(x)\,h$. This agrees with the Seiberg-Witten
differential obtained from geometric engineering provided $h= -{1\over 2\pi i}
d\log(z)$. We proceed to check that this is indeed the case.

The holomorphic 1-form $-{1\over 2\pi i} d\log(z)$ has the same periods over A-cycles as $h$. Both also have
zero periods over the compact B-cycles $C_i = B_i - B_0$. We only need to check non-compact B-cycles.

From the variation with respect to $S$ \delS\ we find:
\eqn\aa{ \sum_{i=1}^n N_i \int_{C_i} \omega_S +
N\int_{-\tilde{\infty}}^{ +\infty}\!\!\omega_S
+ k_q\int_{+\tilde{\infty}}^{-\tilde{\infty}} \!\!\omega_S
+ \int_{ -\tilde{m}}^{+\tilde{\infty}}\!\!\omega_S + 4\pi i\tau = 0 .}
where we put
\eqn\aa{ \omega_S = 2 (N
g_N\del_{b_{N}}+g_{-1}\del_{b_{-1}})\lambda, }
and $\tau$ is given
by \hnoncompact.
Since $\sum_{i=1}^n N_i \int_{C_i} \del_{b_j}\lambda = 0$ for $j=0,\ldots , n-1$ and $\del_S \lambda$ is holomorphic
away from the asymptotic regions,
we get $\sum_{i=1}^n N_i \int_{C_i} \omega_S =0$. Next we introduce
the cut-offs $x\sim (R\Lambda_+)^2, x\sim (R\tilde{\Lambda}_+)^2$ for $+\infty,+\tilde{\infty}$
and $x\sim 1/(R\tilde{\Lambda}_-)^2$ for $-\tilde{\infty}$. As mentioned before we can interpret
these as masses of additional hypermultiplets, eg. $R\Lambda_+ \sim e^{RM_+/2}$. Then
\eqn\aa{
2\pi i\tau = -\log((R\Lambda_+)^{N} (R\tilde{\Lambda}_+)^{k-1} (R\tilde{\Lambda}_-)^{N-k})  + C +
{\cal O}(1/R\Lambda_+,1/R\tilde{\Lambda}_{\pm}) }
for some constant $C$ independent of the lambdas. We have to match this
with \hnoncompact. Unfortunately we have not been able to compute the subleading terms,
so we will only check that the leading divergent pieces agree. Using $h= -{1\over 2\pi i}
d\log(z)$ with z given in \zdef, we get
\eqn\aa{
2\pi i\tau = {1\over 2}\log\left(
{\gamma^2 e^{-Rm}}\over(-16p_0^2)(R\Lambda_+)^{2N}(R\tilde{\Lambda}_+)^{2k-2}(R\tilde{\Lambda}_-)^{2N-2k}
\right)
}
which indeed has the same divergent behaviour.

To get agreement with \SWcurve\ we want to identify $\gamma \sim 4 (R\Lambda)^{2N-1}
e^{Rm/2}$. Then we identify $(R\Lambda_+)^{N}(R\tilde{\Lambda}_+)^{k-1} (R\tilde{\Lambda}_-)^{N-k}$ with the gauge theory
cut-off $(R\Lambda_0)^{2N-1}$ (the factor of $-p_0^2$ presumably gets taken care of by a counterterm
$\pi i S + \log(p_0^2)S$ to the superpotential; see \CachazoYC\ for similar terms in four dimensions).

\bigskip
\noindent
{\sl Acknowledgements:}

It is a pleasure to thank F.~Cachazo and C.~Vafa for useful
discussions, and C.~Vafa for suggesting the project. I would also like to thank
Caltech for hospitality. This work was
supported in part by grants NSF-9802709 and NSF-0243680.

\listrefs

\bye